\documentstyle[prl,aps,multicol,epsf,amsmath]{revtex}

\begin{document}

\title{Learning to coordinate in a complex and non-stationary world}

\author{M. Marsili$^1$, R. Mulet$^2$\cite{MULET},
F. Ricci-Tersenghi$^2$ and R. Zecchina$^2$}

\address{
$^1$ Istituto Nazionale per la Fisica della Materia (INFM), Unit\`a di
Trieste--SISSA, I-34014 Trieste, Italy\\
$^2$ The Abdus Salam International Center for Theoretical Physics,
Condensed Matter Group\\
Strada Costiera 11, P.O. Box 586, I-34100 Trieste, Italy
}

\date{\today}  

\maketitle 

\begin{abstract}
We study analytically and by computer simulations a complex system of
adaptive agents with finite memory.  Borrowing the framework of the
Minority Game and using the replica formalism we show the existence of
an {\em equilibrium} phase transition as a function of the ratio
between the memory $\lambda$ and the learning rates $\Gamma$ of the
agents.  We show that, starting from a random configuration, a {\em
dynamic} phase transition also exists, which prevents the system from
reaching any Nash equilibria.  Furthermore, in a non-stationary
environment, we show by numerical simulations that agents with
infinite memory play worst than others with less memory and that the
dynamic transition naturally arises independently from the initial
conditions.
\end{abstract}


\vspace{-.5cm}
\begin{multicols}{2}
\narrowtext           

Social interactions pose many coordination problems to individuals.
Generally social agents face problems of sharing and distributing
limited resources in an optimal way. Examples range from the use of
public roads and the Internet, to exchanging what we produce with what
we consume. A solution to problem of this kind invokes the
intervention of a public authority who finds the social optimum and
imposes or suggests the optimal behavior to agents.  While such a
solution may be easy to find, its implementation may be difficult to
enforce in practical situations.

Self-enforcing solutions -- where agents achieve optimal allocation of
resources while pursuing their self-interests, without explicit
communication or agreement with others -- are of great practical
importance. Competitive markets are the prototypical example of such a
solution: With everybody maximizing his own profit and no one really
caring for global optimality, competitive markets perform the
remarkable task of leading to system wide optimality.

Micro-economics and Game Theory have gone quite far in explaining what
equilibria can one expect in social interactions.  However most of
these studies deal with unrealistic cases with either few players or
with many, but identical, agents. Secondly the analysis is restricted
to the equilibria which deductively rational players would agree
upon. Such an approach seems unrealistic in cases involving many
individuals with different goals and characteristics.  The
computational complexity required by deductive rationality may easily
go far beyond the capabilities of agents.  Inductive thinking, as
suggested by Arthur~\cite{Arthur}, may be a more suited model of how
real people behave. A growing effort has indeed been put in recent
years in understanding under what conditions bounded inductively
rational agents may reach optimal
outcomes~\cite{Learn,CamererHo}. Several learning rules have been
found to lead to optimal outcomes when a single agent ``plays''
against nature~\cite{Rust}. Similar results hold for games with few
players, even though non-trivial dynamical effects can also
arise~\cite{Learn}.

In this letter we address the problem of how many heterogeneous
adaptive agents learn to coordinate in a complex, eventually
non-stationary, world.  We draw inspiration from recent work on the
Minority Game~\cite{CZ1}, in order to model a typical situation where
a large number of agents pursue different individual goals, using a
certain number of distributed resources.  Optimal use of resources
becomes then a complex coordination problem.

We focus on agents with finite memory and finite learning rates.  We
find that, when agents need to ``learn'' collectively a fixed
structure of interactions, they can attain a close to optimal
coordination, provided that their memory extends far enough into the
past. As the memory decreases, the system undergoes a phase transition
to a state where agents are unable to learn and play in a random way.

More interestingly we find situations where the agents are unable to
coordinate and to converge to a Nash equilibrium.  Thus the game ends
in a stationary regime with no cooperation.  This is a completely
dynamical effect which prevents the system from a proper convergence
to equilibrium and makes useless the standard analysis based on Nash
equilibria.  This is a further clear evidence of the relevance of
tools and ideas of statistical mechanics in the study of complex
socio-economic systems, indeed dynamical transitions are very well
known in statistical mechanics~\cite{BCKM}.

The model we study is closely related to the Minority Game (MG). The
reason for this choice is that this allows us to benefit from the
detailed understanding which has been recently uncovered by the
statistical mechanics approach~\cite{CMZ,dMM}.  On one hand we can
make reference to exact results, on the other we can extend our
understanding of this keystone model of complex adaptive systems.

The model is precisely defined as follows~\cite{CZ1,CMZ}: Agents live
in a world which can be in one of $P$ states, labelled by an integer
$\mu= 1, \dots, P$. Each agent $i=1, \dots, N$ can choose between two
personal strategies, labeled by a spin variable $s_i$, which prescribe
an action $a^{\mu}_{s_i,i}$ for each state $\mu$.  These actions are
drawn from a bimodal distribution for all $i,s$ and $ \mu$, such that
there are two possible actions, do something $(a^{\mu}_{s_i,i}=1)$ or
do the opposite $(a^{\mu}_{s_i,i}=-1)$.

The payoff received by an agent who plays strategy $s_i$, while her
opponents take strategies $s_{-i}=\{s_j,\forall j\neq i\}$, is, in the
state $\mu$,
\begin{equation}
u_i^{\mu}(s_i,s_{-i}) = -a^{\mu}_{s_i,i} A^\mu \quad ,
\label{eq:payoffs_definition}
\end{equation}
where $A^\mu = \sum_j a^{\mu}_{s_j,j}$. The total payoff to agents is
always negative: The majority of agents receives a negative payoff
whereas only the minority of them gain.

The game is repeated many times; the state $\mu$ is drawn from a
uniform distribution $\rho^{\mu}=1/P$ at each time and agents try to
estimate, on the basis of past observations, which of their strategies
is the best one. More precisely, if $s_i(t)$ is the strategy played by
agent $i$ at time $t$, we assume as in~\cite{Cavagna} that
\begin{equation}
{\rm Prob}[s_i(t)=s] \propto \exp\left[ \Gamma U_{s,i}(t) \right]
\quad ,
\label{eq:probability_definition}
\end{equation}
where $U_{s,i}(t)$ is the {\em score} of strategy $s$ at time $t$ and
$\Gamma$ is a positive constant \cite{McFadden}. Each agent monitors
the scores $U_{s,i}(t)$ of each of her strategies $s$ by
\begin{equation}
U_{s,i}(t+1) = (1- \lambda/P) U_{s,i}(t) + u_i^\mu[s,s_{-i}(t)]/P
\quad ,
\label{eq:evolution}
\end{equation}
where the last term is the payoff agent $i$ would have received if she
had played strategy $s$ at time $t$ -- see
Eq.~(\ref{eq:payoffs_definition}) -- against the strategies
$s_{-i}(t)=\{s_j(t),~\forall j\neq i\}$ played by her opponents at
that time.

In words, Eqs.~(\ref{eq:probability_definition},\ref{eq:evolution})
model agents who play more likely strategies which have performed
better in the past.
Eqs.~(\ref{eq:probability_definition},\ref{eq:evolution}) belong to a
class of learning models which has received much attention
recently~\cite{CamererHo}.

The relevant parameter~\cite{Savit} is the ratio $\alpha=P/N$ between
the ``information complexity'' $P$ and the number of agents, and the
key quantity we shall look at is the global efficiency defined as
$\sigma^2=\langle{A^2}\rangle$.

This model differs from the MG~\cite{CZ1} for two important aspects:
First agents compute correctly the payoff for strategies $s\ne s_i(t)$
which they did not play. In the MG agents only account for the
explicit dependence of $u_i^\mu$ on $s$ which arises from
$a_{s,i}^\mu$ -- see Eq.~(\ref{eq:payoffs_definition}) -- whereas they
neglect the fact that if they had taken a different decision also
$A^\mu$ would have changed. This seems reasonable at first sight
because $A^\mu$ is an aggregate quantity and its dependence on each
individual agent is weak.  A more careful analysis \cite{CMZ,dMM}
however shows that if agents properly account for their impact on
$A^\mu$ as in Eq.~(\ref{eq:evolution}) a radically different scenario
arises: Rather than converging to an unique stationary state as in the
MG, the dynamics (with $\lambda=0$) converges to one of exponentially
many (in $N$) states -- which are Nash equilibria \cite{Nash} --
characterized by an optimal coordination.  This change emerges in the
statistical mechanics approach with the breakdown of replica symmetry
(RS): While the Minority Game is described by a replica symmetric
theory, Nash equilibria are described by a full replica symmetry
broken (RSB) phase~\cite{dMM}. Our aim is precisely that of studying
the coordination of adaptive agents in a complex world with
exponentially many optimal states (Nash equilibria).

The second key feature is that previous work has only explored the
dynamics of learning with an infinite memory -- i.e.\ with $\lambda=0$
in Eq.~(\ref{eq:evolution}) -- and for a fixed structure of
interactions -- i.e.\ with fixed (quenched) disorder $a_{s,i}^\mu$.
Our goal is to clarify the role of different time-scales involved in
the learning dynamics. We shall first study the case where the
structure of interactions is fixed -- which corresponds to
$a_{s,i}^\mu$ being the usual quenched disorder -- and then move to
the more realistic case where the structure of interactions changes
over long time-scales.

Following the lines of reasoning of Refs.~\cite{CMZ,errata}, we
introduce a continuum time $\tau=\Gamma t/P$ and variables
$y_i(\tau)=\Gamma [U_{+,i}(t)-U_{-,i}(t)]/2$ in terms of which the
dynamics reads
\begin{equation}
\frac{dy_i}{d\tau} = -\frac{\lambda}{\Gamma} y_i -h_i
-\sum_{j \neq i}J_{i,j} \tanh(y_j) +\eta_i(\tau) \quad ,
\label{lang}
\end{equation}
\begin{eqnarray*}
h_i&=&\frac{1}{P}\sum_{\mu=1}^P\sum_{j=1}^N 
\frac{a_{+,i}^\mu-a_{-,i}^\mu}{2}\frac{a_{+,j}^\mu+a_{-,j}^\mu}{2}
\quad ,\\
J_{i,j}&=&\frac{1}{P}\sum_{\mu=1}^P
\frac{a_{+,i}^\mu-a_{-,i}^\mu}{2}\frac{a_{+,j}^\mu-a_{-,j}^\mu}{2}
\quad ,
\end{eqnarray*}
with $\eta_i(\tau)$ a white noise with zero mean and correlations
\[
\langle \eta_i(\tau)\eta_j(\tau') \rangle \simeq \frac{\Gamma
\sigma^2}{\alpha N} \delta_{i,j}\delta(\tau-\tau') \quad .
\]
Refs.~\cite{CMZ,errata} have shown that, for $\lambda=0$, the
stationary states of this dynamics are related to the local minima of
\[
\sigma^2 = H_0 +2\sum_i h_i m_i +\sum_{j\neq i}J_{i,j}m_i m_j \quad ,
\]
where $H_0$ is a constant and $m_i=\langle\tanh(y_i)\rangle$.  These
states are also Nash equilibria~\cite{Nash}, which means that agents
achieve an optimal coordination.  Since $\sigma^2$ takes its minima
for $m_i=\pm 1$ -- which correspond to $y_i\to\pm\infty$ -- the
stochastic force $\eta_i(t)$ is irrelevant in the late stages of the
dynamics, which is dominated by the deterministic drift towards the
Nash equilibrium.

For $\lambda/\Gamma>0$ we expect the stochastic force $\eta_i(\tau)$,
whose strength is itself proportional to $\sigma^2$, to compete with
the deterministic drift. Indeed the distribution of $y_i$ will be
cutoff for $|y_i|\gg \Gamma/\lambda$: For small $\lambda$ we expect
that $\langle\tanh(y_i)\rangle$ is close to the values
$m_i^{(\infty)}$ which minimize $\sigma^2$, and a spread in the
distribution of $y_i$ around its average which is maintained by the
stochastic force.  When $\lambda$ increases we expect a transition to
a phase where agents are unable to coordinate because their memory is
too short for learning correctly the interaction structure: The
dynamics is dominated by the stochastic force $\eta_i$, which is made
even stronger by the fact that $\sigma^2/N\simeq 1$ is much larger
than in the coordinated state. This transition is captured by the
statistical mechanics approach of Ref.~\cite{CMZ}. Neglecting
stochastic fluctuations induced by $\eta_i$, which is legitimate only
for $\Gamma\ll 1$, one can easily prove, following Ref.~\cite{CMZ},
that $m_i^{(\lambda)}=\langle\tanh y_i\rangle$ are given by the
solution of the minimization of the function
\begin{equation}
H = \sigma^2 + \frac{\lambda}{\Gamma} \sum_i \left[ \log(1-m_i^2) + 2
m_i \tanh^{-1}(m_i) \right] \; .
\label{eq:definition_of_H}
\end{equation}
In order to study the ground state properties of $H$ we follow the
same steps of Ref.~\cite{CMZ}: We introduce an inverse temperature
$\beta$, we compute the partition function and the free energy per
agent and then we take averages over the disordered variables
$a^{\mu}_{s,i}$ with the replica method~\cite{MPV}.  The free energy,
within the RS Ansatz, reads
\begin{gather}
f(q,r,Q,R) = \frac{\alpha}{\beta} \ln\left[ 1 +
\frac{\beta(Q-q)}{\alpha}\right ] + \frac{\alpha}{2}
\frac{1+q}{\alpha+\beta(Q-q)} \nonumber \\
+ \frac{1-Q}{2} - \frac{1}{\beta} \langle \ln \int_{-1}^{1} \! \! {\rm
d}m\, e^{-\beta V_z(m)} \rangle + \frac{\alpha \beta}{2} (RQ-rq) \, ,
\label{eq:free_energy}
\end{gather}
where $Q=\frac{1}{N}\sum_i (m_i)^2$ and $q = \langle m_i^a m_i^b
\rangle$ with $a \neq b$ labelling different replicas of the systems;
$R$ and $r$ arise as Lagrange multipliers and $V_z(m)=-\sqrt{\alpha r}
mz +\frac{\alpha \beta}{2}(r-R)m^2 +\frac{\lambda}{\Gamma}[\log(1-m^2)
+2 m \tanh^{-1}(m)]$.  The ground state properties of $H$ are obtained
solving the saddle point equations~\cite{MPV} in the limit $\beta
\rightarrow \infty$.

\begin{figure}
\epsfxsize=0.95\columnwidth
\epsffile{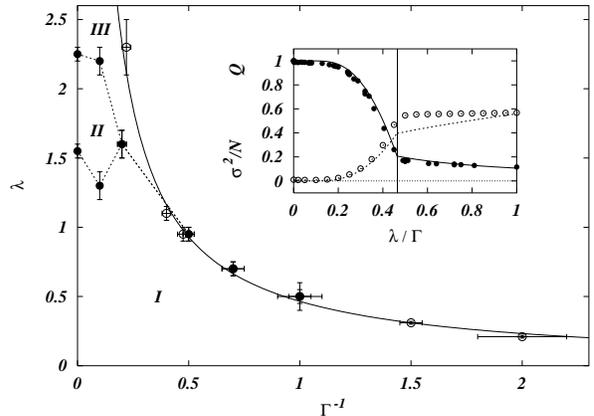}
\caption{Phase diagram: static ($\circ$) and dynamic ($\bullet$)
critical lines obtained from the simulation.  The full line represents
the RS critical line.  The dashed lines are guide to the eyes.  Inset:
$Q$ ($\bullet$) and $\sigma^2/N$ ($\circ$) as a function of
$\lambda/\Gamma$ obtained for the simulation. The lines represent the
RS solution.}
\label{fig1}
\end{figure}

In the inset of Fig.~\ref{fig1} we compare the analytical predictions
for $\sigma^2$ and $Q$ with simulations results.  We focus on small
$\alpha$ (i.e.\ $\alpha=0.1$) where the effects we wish to discuss are
more evident.  Little discrepancies between numerical data and
analytical curves are maybe due to RSB effects.  Note that a phase
transition occurs at $\lambda_c \simeq 0.46 \Gamma$ where both
$\sigma^2$ and $Q$ change their analytical behaviour.  We have studied
this equilibrium phase transition in the $(\lambda,1/\Gamma)$ plane,
confirming the critical line $\lambda_c=0.46 \Gamma$: Open symbols in
Fig.~\ref{fig1} refer to a {\em static} experiment where we let the
system equilibrate to a Nash equilibrium for $\lambda=0$ and then
we move it slowly along lines $\lambda\Gamma={\rm const}$.

The situation changes when the system starts from scratch
[$U_{s,i}(0)\!=\!0 \; \forall\{s,i\}$] in each run.  Depending on
$\lambda$ and $\Gamma$, the dynamics may lead the system to a
stationary regime (different from the static one) which is
characterized by larger fluctuations (i.e.\ larger $\sigma^2$).  These
dynamical effects make the phase diagram more complex in the
$\lambda<\lambda_c$ region (see Fig.~\ref{fig1}): In I the system
always relaxes to the static equilibrium, in II it sometimes converges
to equilibrium and sometimes get trapped in the metastable regime with
large fluctuations, while in III it never reaches equilibrium. The
presence of this dynamical transition implies that the analysis in
terms of Nash equilibria is no longer enough to predict the collective
behavior of the system in a large part of the phase diagram, i.e.\ for
high learning rates and short memory.

When the external world is non-stationary, i.e.\ changes with time,
the adaptation task becomes still harder.  We mimic the external world
modification as follows: Every $\tau$ time steps a randomly chosen
state of the world is removed and a new one replaces it (in order to
keep $P$ constant).  Actually we randomly choose a $\mu$ index and we
re-extract the strategies $a^\mu_{s,i}$ for all $i$ and $s$.

Here we focus on the results of the simulations done with $\tau=10^3$,
$\Gamma=\infty$, $NP=10^4$ and many $\lambda$ values.  The results do
not dependent on the initial conditions.

\begin{figure}
\epsfxsize=0.95\columnwidth
\epsffile{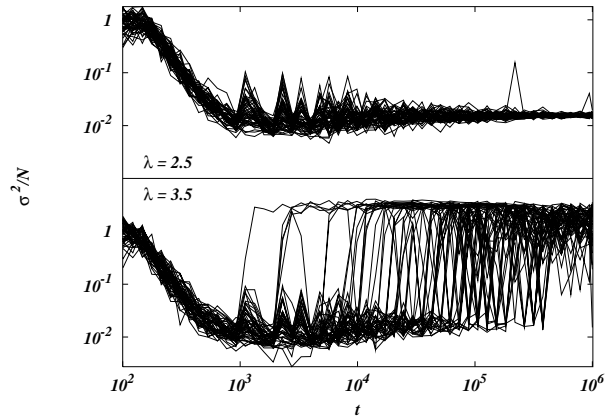}
\caption{In a non-stationary world ($\tau=10^3$) the evolution of
$\sigma^2/N$ with simulation time for 50 different samples and two
values of $\lambda$ ($NP=10^4$, $\alpha=0.1$ and $\Gamma=\infty$).}
\label{fig2}
\end{figure}    

In the upper panel of Fig.~\ref{fig2} we show the relaxation of
$\sigma^2/N$ for $\lambda=2.5$: As expected, it starts from 1 and
converges to its equilibrium value.  Note that $\tau=10^3$ has been
chosen in order to allow the system to reach a cooperative behaviour
before the world starts changing.  For this value of $\lambda$ the
system is robust with respect to changes of the world: Apart from
occasional excursions to states with large $\sigma^2$, agents are able
to adapt themselves to the evolving interaction structure.

In the lower panel we present the evolution of $\sigma^2/N$ for
$\lambda=3.5$ (i.e.\ with shorter memory) in 50 different samples.
The behaviour is now completely different: After having reached a low
value of $\sigma^2/N$ (cooperation) the system undergoes a sharp
transition and $\sigma^2/N$ jumps to a high value. The players are no
longer able to adapt to the changing world and they start playing in a
wrong way. Occasionally agents may achieve a good coordination with
small $\sigma^2$, but they eventually always go back to uncoordinated
states with large $\sigma^2$.

\begin{figure}
\epsfxsize=0.95\columnwidth
\epsffile{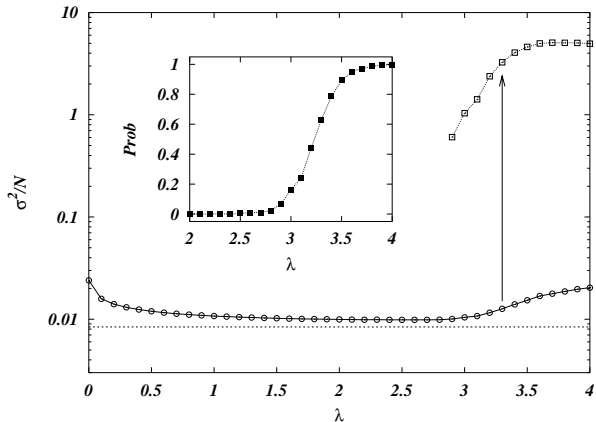}
\caption{Average low ($\circ$) and high ($\Box$) $\sigma^2/N$ as a
function of $\lambda$ ($NP=10^4$, $\alpha=0.1$, $\Gamma=\infty$ and
$\tau=10^3$).  The arrow indicates a transition from the cooperative
to the non-cooperative regime.  The horizontal dotted line is the
$\sigma^2/N$ value with fixed world ($\tau=\infty$).  Inset:
Probability of being in a non-cooperative regime as a function of
$\lambda$.}
\label{fig3}
\end{figure}

For large times, the instantaneous values of $\sigma^2/N$ have a
roughly bimodal distribution: They are either low ($\sim 10^{-2}$) or
high ($\sim 1$).  In Fig.~\ref{fig3} we plot the average of the low
($\circ$) and of the high ($\Box$) values (these averages can be
defined in an unambiguous way thanks to the gap between low and high
$\sigma^2$ values).  In the inset we report the fraction of samples
that spend the last decade in the high $\sigma^2$ regime. In a whole
intermediate range around $\lambda_c\approx 3.3$ we find that
coordinated states with small $\sigma^2$ coexist with wildly
fluctuating states ($\sigma^2>1$).

Is worth noticing some facts in Fig.~\ref{fig3}.  The minimum of
$\sigma^2$, corresponding to the best cooperation, is no longer
located in $\lambda=0$ (i.e.\ infinite memory).  In other words, in a
non-stationary environment the agents play better with a {\em finite}
memory, which allows them to take decision based more on the recent
past rather than on the far past.  The minimum they can attain is very
near to the $\sigma^2/N$ value in an unchanging world (shown with a
horizontal line in Fig.~\ref{fig3}). The second remarkable fact is
that the transition from a coordinated state to a high $\sigma^2$
regime when $\lambda$ increases -- which was continuous in a fixed
world -- shows features of first order transitions such as
discontinuities and phase coexistence.

In conclusion, we have extended the replica solution of the Minority
Game to the case where agents have finite memory and finite learning
rates.  We have proven that a phase transition between phases with low
and high $\sigma^2$ exists as a function of $\lambda/\Gamma$.  We have
also shown, by means of computer simulations, that a dynamical phase
transition exists for high values of $\lambda$ (short memories), and
that this dynamic phase transition is responsible for a
non-cooperative behaviour of agents.  Furthermore we have shown, by
numerical simulation, that when the structure of the interactions is
non-stationary, agents with infinite memory behave worst than agents
with a finite memory.  Under these conditions we recover again a
scenario where agents with too short memory display a first order
transition from a cooperative to a non-cooperative phase.

\end{multicols}  


\vspace{-.3cm}
\begin{thebibliography}{99}
\vspace{-1.5cm}

\bibitem[\dag]{MULET} Permanent address: Supercond. Lab.,
Fac. F\'{\i}sica - IMRE, Univ. Havana, CP 10400, La Habana, Cuba.

\bibitem{Arthur} W.B. Arthur, Am. Econ. Assoc. Papers Proc. {\bf 84},
406 (1994).

\bibitem{Learn} D. Fudenberg and D.K. Levine, {\it The theory of
learning in games} (MIT Press, 1998).

\bibitem{CamererHo} C. Camerer and T.-H. Ho, Econometrica {\bf 67},
827 (1999).

\bibitem{Rust} A. Rustichini, Games and Economic Behavior {\bf 29},
244 (1999).

\bibitem{CZ1} D. Challet and Zhang Y.-C., Physica A {\bf 246}, 407
(1997).

\bibitem{BCKM} J.P. Bouchaud, L.F. Cugliandolo, J. Kurchan and
M. M\'ezard, in {\it Spin Glasses and Random Fields}, A.P. Young
ed. (World Scientific, Singapore, 1998).

\bibitem{CMZ} D. Challet, M. Marsili and R. Zecchina,
Phys. Rev. Lett. {\bf 84}, 1824 (2000); M. Marsili, D. Challet and
R. Zecchina, Physica A {\bf 280}, 522 (2000).

\bibitem{dMM} A. De Martino and M. Marsili, J. Phys. A {\bf 34}, 2525
(2001).

\bibitem{Cavagna} A. Cavagna, J.P. Garrahan, I. Giardina, and
D. Sherrington, Phys. Rev. Lett. {\bf 83}, 4429 (1999).

\bibitem{McFadden} The probabilistic nature of agent's choice does not
necessarily imply that agents randomize their behavior on purpose.  Mc
Fadden [Ann. Econ. Soc. Measurement, {\bf 5}, 363 (1976)] has indeed
shown that Eq.~(\ref{eq:probability_definition}) models individuals
who maximize an ``utility'' which has an implicit random idiosyncratic
part.  The constant $\Gamma$ is then the relative weight which agents
assign to the empirical evidence accumulated in $U_{s,i}$ with respect
to random idiosyncratic shocks: If $\Gamma \rightarrow \infty$ agents
always play their best strategy according to the scores, while if
$\Gamma$ decreases agents take less into account past performances.

\bibitem{Savit} R. Savit, R. Manuca and R. Riolo,
Phys. Rev. Lett. {\bf 82}, 2203 (1999).

\bibitem{Nash} $\{s_i^*,~\forall i\}$ is a Nash equilibrium if for all
$i$, $u_i(s,s_{-i}^*)\le u_i(s_i^*,s_{-i}^*)$ holds $\forall s$. See
D. Fudenberg and J. Tirole, {\it Game Theory} (MIT Press, 1991).

\bibitem{errata} M. Marsili and D. Challet, e-print cond-mat/0102257.

\bibitem{MPV} M. M\'ezard, G. Parisi, M. A. Virasoro, {\it Spin glass
theory and beyond} (World Scientific, Singapore, 1987).

\end{thebibliography}
\end{document}